# Slip Effects in Compressible Turbulent Channel Flow

P. A. Skovorodko

*Kutateladze Institute of Thermophysics, 630090, Novosibirsk, Russia*

**Abstract.** The direct numerical simulation of compressible fully developed turbulent Couette flow between two parallel plates with temperature $T_w$ moving with velocities $\pm U_w$ was performed. The algorithm was tested on well known numerical solution for incompressible Poiseuille channel flow and found to provide its well description. The slip effects in studied flow are found to be negligibly small at the values of accommodation coefficients $\alpha_u$ and $\alpha_T$ of the order of unity. The considerable increase of mean temperature with decreasing the accommodation coefficient $\alpha_T$ for fixed value of $\alpha_u = 1$ was discovered. The effect may be important in the problems of heat exchange in compressible turbulent boundary layer for some combinations of flowing gas, surface and adsorbing gas.

## INTRODUCTION

Despite significant advances in computer technology the possibility of direct simulation of practically important classes of turbulent flows is very limited. The planar Poiseuille and Couette flows are widely used for testing algorithms for the direct simulation of turbulent flows, since, in spite of the simple geometry, they represent an example of the flow, preserving the main properties of a turbulent boundary layer and the turbulent flow itself [1 - 3]. At these flows it is convenient to test various approximate models of turbulence.

The simulation of compressible turbulent flows is more difficult since besides the Reynolds number, the only determining parameter for incompressible flow, the considered problem contains at least eight additional determining parameters even for the flow of monocomponent perfect gas. The effect of two of them, namely, the accommodation coefficients for tangential velocity and for temperature will be analyzed here.

One of the most probable effects of rarefaction on compressible turbulent channel flow may be connected with interaction between the flowing gas molecules and the surfaces of the channel. The first approximation for taking this interaction into account leads to velocity slip and temperature jump boundary conditions on the surface. In spite of high Reynolds number typical of turbulent flows, the negligibly small influence of the slip effects on the flow field is not evident taking into account high gradients of velocity and temperature near the surface. To estimate the value of the discussed effects the DNS of compressible fully developed turbulent Couette flow between two parallel plates with temperature $T_w$ moving with velocities $\pm U_w$ was performed.

In the papers devoted to the direct simulation of turbulent planar Poiseuille and Couette flows, the grid with highly non-uniform partitioning of the coordinates *y*, normal to the surface, is usually applied – the cell size near the surface can be smaller than that in the middle part of the flow in more than 10 times [1 - 3] . Meanwhile, the distribution of mean parameters in the laminar sublayer is close to linear and the level of pulsation of the main parameters of the flow is small, so the need for a substantial thickening of the finite-difference grid in the vicinity of the surface to correct description of the flow is not obvious. In the other words, it is interesting to clarify the question of errors of flow description by using a grid with a sufficiently large size of the cell near the surface, for example, using a grid with a uniform partition of the coordinate normal to the surface. The situation is similar with the necessary sizes of the computational domain, which is a parallelepiped, along *x* and *z* coordinates – is not clear how these dimensions can be reduced without significant distortions of the basic characteristics of the flow.

In the first part of this paper, a direct simulation of turbulent planar Poiseuille and Couette flows of an incompressible fluid using a finite-difference grid with uniform partition of the coordinate normal to the surface in the computational domain rather smaller than usually used in the literature is performed. The main objective of this part – to determine the accuracy of description of the main characteristics of the flow on the used finite-difference grid. The study is a necessary step before the direct simulation of turbulent flow of compressible gas, with no-slip or slip boundary conditions.

## PROBLEM FORMULATION

Direct simulation of turbulent planar Poiseuille and Couette flows of incompressible fluid is performed in the frames of time-dependent Navier-Stokes equations using the method of artificial compressibility proposed by N.N. Yanenko [4]:

$$a\frac{\partial p}{\partial t} + \frac{\partial u_i}{\partial x_i} = 0, \tag{1}$$

$$\frac{\partial u_i}{\partial t} + u_j \frac{\partial u_i}{\partial x_j} + \frac{\partial p}{\partial x_i} = \frac{1}{\text{Re}} \nabla^2 u_i. \tag{2}$$

The system of equations (1, 2) is written in dimensionless form, with the spatial coordinates being expressed in the units of half-width of the channel $\delta$, the velocity component – in terms of friction velocity $u_\tau = \sqrt{\tau_w/\rho}$ ($\tau_w$ is the shear stress, $\rho$ is the fluid density), and the pressure $p$ – in units $\rho u_\tau^2$. The Reynolds number $\text{Re}$ is equal to $u_\tau \delta / \nu$ where $\nu$ is the kinematic viscosity.

Dimensionless parameter $a$ in the continuity equation (1) must be chosen for reasons of minimizing the perturbations introduced to the solution by the term describing an artificial compressibility. From a comparison of equation (1) with the continuity equation for the isothermal flow of a compressible gas is easy to establish that the parameter $a$ is equal to the square of ratio of friction velocity to the sound velocity, in the other words, the square of the corresponding Mach number (M). This allowed us to select a value $a = 10^{-4}$ that corresponds to M = 0.01, which was used in most of the calculations. Test calculations of some variants with $a = 10^{-3}$ did not reveal any significant impact of a parameter $a$ on the results that allows us to hope for a little difference of the obtained results from the solution corresponding to the value $a = 0$.

Below the coordinate along the flow direction will be denoted by $x$, the coordinate normal to the surface – by $y$, and the coordinate orthogonal to the two mentioned – by $z$. The velocity components along the coordinates $x$, $y$ and $z$ will be denoted by $u$, $v$ and $w$, respectively. The computational domain for both problems was a parallelepiped with dimensions $L_x$, $L_y = 2\delta$ and $L_z$ along the corresponding coordinates. All the reported below results were obtained for the region with dimensions $L_x = 1.6\pi\delta$, $L_z = 2\delta$. The uniform grid with 80 cells along each direction was applied with the total number of cells in the calculation being equal to 512000.

The finite-difference equations were written in conservative form on a staggered grid, when the pressure is defined in the center of the cell while the components of velocity – in the middle of corresponding borders of the cell [5]. For fully developed turbulence considered the periodic boundary conditions on corresponding boundaries of the computational domain were set. The velocity components on a solid surface were prescribed to be zero and there is no need to prescribe the pressure here. For the problem of Poiseuille flow periodic conditions for the pressure at the boundaries $x = 0$ and $x = L_x/\delta$ were set taking into account the pressure drop $\Delta p$, which is equal to $L_x/\delta$ in adopted normalization. For the problem of Couette flow the velocity component $u = -U_w$ at $y = 0$ and $u = U_w$ at $y = L_y/\delta$.

The flows were simulated by explicit method, the parameters on the new time layer were determined iteratively with successive refinement of them at the intermediate time level, with corresponding number of iterations being equal to 4. As the result, the order of approximation over time is close to second. The second order approximation over spatial variables is automatically ensured by the properties of the staggered grid.

For the Poiseuille flow the regime with Reynolds number Re = 180 is considered for which the literature contains the results of calculations obtained by the spectral method [1, 6]. The initial distribution of the parameters in the flow field was obtained by calculating the temporal evolution of the flow from laminar distribution of parameters subjected to random perturbations. This evolution can be traced as long as the distribution of parameters, averaged over a certain time interval, ceased to depend on time. For the Couette flow, the initial distribution of the parameters was obtained based on the distributions for the Poiseuille flow with a corresponding adjustment of the fields of pressure and velocity components. The calculations were performed with a time step $dt = 1.366 \cdot 10^{-4}$. The number of time steps, which was conducted by averaging the flow parameters, was $3 \cdot 10^5$ for the Poiseuille flow and $2.5 \cdot 10^5$ for the Couette flow.

In the simulation of compressible Couette flow we use the full set of unsteady 3D Navier-Stokes equations written in conservative form:

$$\frac{\partial \rho}{\partial t} + \frac{\partial \rho u_i}{\partial x_i} = 0, \tag{3}$$

$$\frac{\partial \rho u_i}{\partial t} + \frac{\partial \rho u_i u_j}{\partial x_j} + \frac{\partial p}{\partial x_i} = \frac{\partial \tau_{ij}}{\partial x_j}, \tag{4}$$

$$\frac{\partial \rho e}{\partial t} + \frac{\partial (\rho e + p) u_i}{\partial x_i} = \frac{\partial u_i \tau_{ij}}{\partial x_j} - \frac{\partial q_i}{\partial x_i}, \tag{5}$$

where

$$\tau_{ij} = \mu \left( \frac{\partial u_i}{\partial x_j} + \frac{\partial u_j}{\partial x_i} - \frac{2}{3} \delta_{ij} \frac{\partial u_k}{\partial x_k} \right) + \mu' \delta_{ij} \frac{\partial u_k}{\partial x_k},$$

$$q_i = -\lambda \frac{\partial T}{\partial x_i},$$

$$e = \frac{RT}{(\kappa - 1)} + \frac{u_i u_i}{2},$$

$$p = \rho RT.$$

Here $\mu$ is the dynamic viscosity, $\mu'$ is the bulk viscosity, $\lambda$ is the heat conductivity, $R$ is the gas constant, $\kappa$ is the specific heat ratio, $e$ is the total internal energy, $T$ is the temperature. The finite-difference approximation of the governing equations was made on the same staggered grid as was used for incompressible flow simulations. In the streamwise and spanwise directions the periodic boundary conditions were applied. On the surfaces of the channel either no-slip or slip boundary conditions were used:

$$\Delta u = U_w - u \mid_w = -\frac{2 - \alpha_u}{\alpha_u} \frac{\mu}{p} \left( \frac{\pi RT}{2} \right)^{1/2} \left( \frac{\partial u}{\partial y} \right)_w, \tag{6}$$

$$\Delta T_w = T_w - T \mid_w = -\frac{2 - \alpha_T}{\alpha_T} \frac{2\kappa}{\Pr(\kappa + 1)} \frac{\mu}{p} \left( \frac{\pi RT}{2} \right)^{1/2} \left( \frac{\partial T}{\partial y} \right)_w, \tag{7}$$

where $\alpha_u$ and $\alpha_T$ are the accommodation coefficients for tangential velocity and for temperature, respectively.

The Couette flow was simulated for perfect gas with specific heats ratio $\kappa = 1.4$. The values of Prandtl number $\Pr = \frac{\mu \kappa R}{(\kappa - 1)\lambda} = 1$, the bulk viscosity $\mu' = 0$, and constant dynamic viscosity $\mu$ were chosen for simplicity. Four regimes corresponding to the values of Mach number, M = $U_w/(\kappa RT_w)^{1/2}$, 0.59, 1, 1.86, and 2.5 were considered. The Reynolds number, $Re_w = \rho_0 U_w \delta/\mu$, was kept constant, $Re_w = 3353$ ($\rho_0$ – the mean density). The accommodation coefficients $\alpha_u$ and $\alpha_T$ entering the expressions for velocity slip (6) and temperature jump (7) were varied in the range $0.01 \div 1$. The non-slip conditions were provided by the same expressions (6), (7) with the values $\alpha_u = 2$, $\alpha_T = 2$. The domain of simulation, the staggered grid and the method of solution of the finite-difference relations were the same as in the simulations of incompressible flows. The initial distribution of the velocity components and the density for compressible Couette flow for the lowest value of Mach number (M = 0.59) was obtained from incompressible flow field while the temperature was set constant and equal to $T_w$. The temporal evolution of the flow was traced as long as the distribution of parameters, averaged over a certain time interval, ceased to depend on time. The initial distributions of the parameters in the domain of simulation for another set of determining parameters was set based on the solution for the regime which is the closest to the considered.

# RESULTS AND DISCUSSION

All the results obtained in [1] are available at [6] in free access that facilitates the process of analysis and comparison with the results of other studies. It also contains the results for higher Reynolds numbers, as set out in [2]. Figures 1 – 4 illustrate the results obtained for incompressible flow.

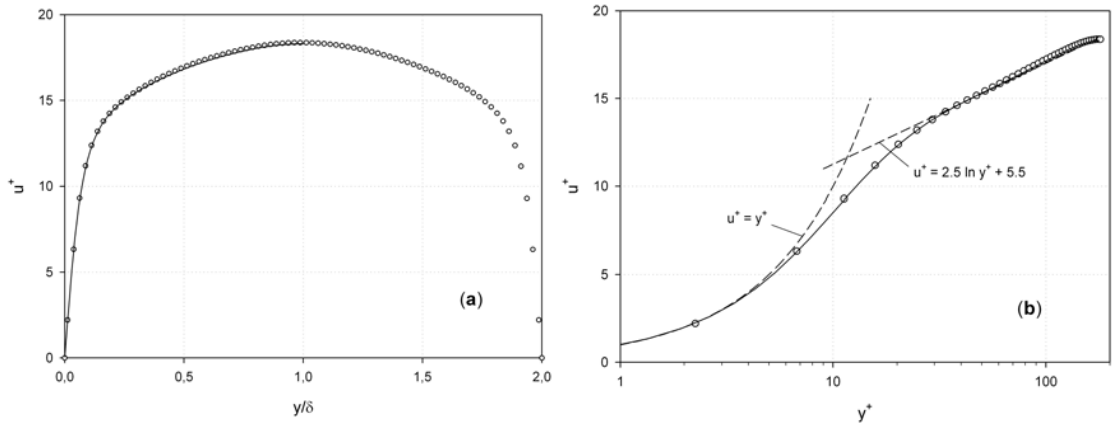

**FIGURE 1.** The mean component $u+$ of velocity for Poiseuille flow at Re = 180 versus $y/\delta$ (a) and universal non-dimensional coordinate $y+ = yu_\tau/v$ (b) (symbols – present study, solid line – the data from [1]).

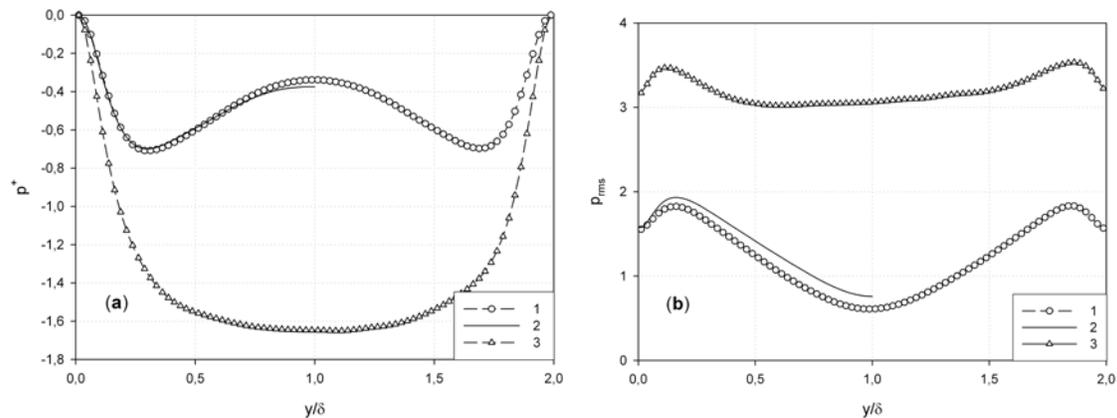

**FIGURE 2.** The profiles of mean pressure (a) and rms pulsations of pressure (b) for Poiseuille flow at Re = 180 (1 – present study, 2 – the data from [1]), as well as for Couette flow at Re = 185 (3).

Figure 1(a) illustrates the profile of mean component $u^+$ of velocity for Poiseuille flow at Re = 180 with both halves of the profile being showed to judge the degree of symmetry of the flow. As can be seen, the present results are in good agreement with the data from [1, 6].

In Figure 1(b) the same profiles are plotted in dependence of universal non-dimensional coordinate $y^+ = yu_\tau/v$ in logarithmic scale. The dashed lines show the distribution typical of the laminar sublayer, as well as a distribution that approximates well the data of [1] in the main part of a turbulent boundary layer. As it is seen from Figure 1(b), the same distribution provides the good approximation of the present results too. The good agreement of present results with the results of [1] takes place, despite the fact that for the grid used here the first calculated point, besides the point on the channel boundary, is located at $y^+ = 2.25$ (see Figure 1(b)).

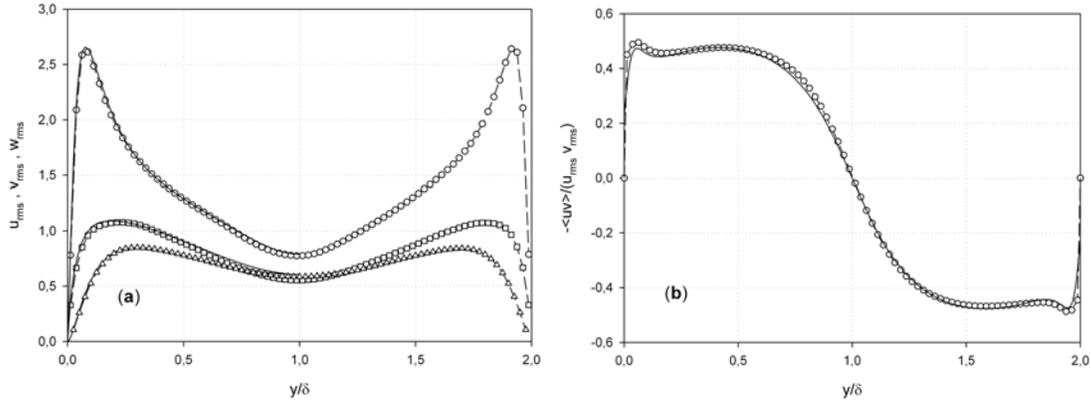

**FIGURE 3.** The profiles of rms pulsations of component $u$ (upper curve), $v$ (lower curve) and $w$ (middle curve) of velocity (a) and the profile of correlation coefficient $-<uv>/(u_{rms}v_{rms})$ (b) for the Poiseuille flow at Re = 180 (the data from [1] are shown by solid lines).

Figure 2(a) illustrates the comparison of mean pressure $p^+$ profile for the Poiseuille flow with the results of [1]. It also shows the profile of mean pressure in the Couette flow at Re = 185. As can be seen, for Poiseuille flow the data of two calculations agree well with each other in almost all of the profile, except a small neighborhood of the plane of symmetry, where between them there is some difference. The nonuniformity of the mean pressure for the Couette flow significantly higher than that for the Poiseuille flow at similar Reynolds numbers. For both the considered flows the maximum mean pressure is reached on the surface of the channel.

Figure 2(b) illustrates the comparison of rms pulsations of pressure ($p_{rms}$) for the same variants as in Figure 2(a). Here, with the same qualitative behavior of the profiles, there is a slight quantitative difference between the results of this work and [1], which is absent on the surface of the channel, but increases to approximately 25% in the plane of symmetry. The reasons for this difference may be caused by a small damping of pressure pulsations by the approach of artificial compressibility used in this paper. Another reason may be due to the difference of 2.5 times in the length of the channel, which is $L_x = 4\pi\delta$ in [1]. The level of pressure pulsations in the Couette flow (curve 3 in Figure 2(b)) is significantly higher than that in the Poiseuille flow at similar Reynolds numbers. Small asymmetry of the pressure pulsation profile for Couette flow (curve 3 in Figure 2(b)) seems to be caused by insufficiently long time interval used for averaging of this very sensitive parameter of the flow.

Figure 3(a) shows the profiles of rms pulsations of velocity components $u$, $v$ and $w$ for the Poiseuille flow, which, apparently, are in good agreement with the data of [1].

Figure 3(b) illustrates the comparison of the profile of correlation coefficient $-<uv>/(u_{rms}v_{rms})$ for the Poiseuille flow with the results of [1]. Again the good agreement between the data of two calculations is observed.

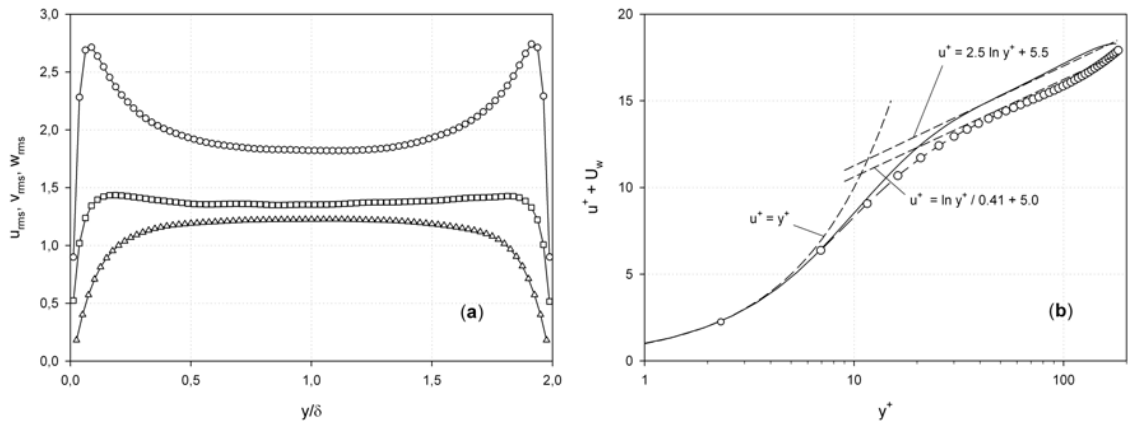

**FIGURE 4.** The profiles of rms pulsations of component $u$ (upper curve), $v$ (lower curve) and $w$ (middle curve) of velocity (a) and the averaged velocity components $u^+ + U_w$ versus universal non-dimensional coordinates $y^+$ (b) for the Couette flow at Re = 185 (symbols – this work, solid line – the results of [1] for the Poiseuille flow at Re = 180).

Other features of the Poiseuille flow, such as the drag coefficient, the ratio of the velocity in the plane of symmetry to the mean flow velocity, the Reynolds stresses distribution are also in good agreement with the results of [1].

For the Couette flow the Reynolds number was found to be Re = 185 for the value of velocity $U_w = 18.14$ (for the Poiseuille flow the mean velocity in the middle of channel was 18.37). The Reynolds number defined by surface velocity $U_w$ at the considered conditions was equal to $Re_w = U_w \delta / \nu = 3353$. That is why the compressible Couette flow was simulated just at this value of $Re_w$.

Figure 4(a) illustrates the profiles of rms pulsations of velocity components $u$, $v$ and $w$ for Couette flow which are similar to those observed for the Poiseuille flow, shown in Figure 3(a).

Figure 4(b) illustrates the profile of mean component $u^+$ of velocity for the Couette flow at Re = 185 in the reference frame associated with the lower surface. The profile is plotted in dependence of universal non-dimensional coordinate $y^+ = y u_\tau / \nu$ in logarithmic scale. The dashed lines show the distribution $u^+ = y^+$ typical of the laminar sublayer, the distribution $u^+ = 2.5 \ln y^+ + 5.5$, which approximates well the data of [1] in the main part of a turbulent boundary layer for the Poiseuille flow (see Figure 1(b)), and the distribution $u^+ = \ln y^+ / 0.41 + 5.0$, used to approximate the distribution in the main part of the boundary layer for the Couette flow at Re = 126 [3], which, in turn, is in agreement with the present results.

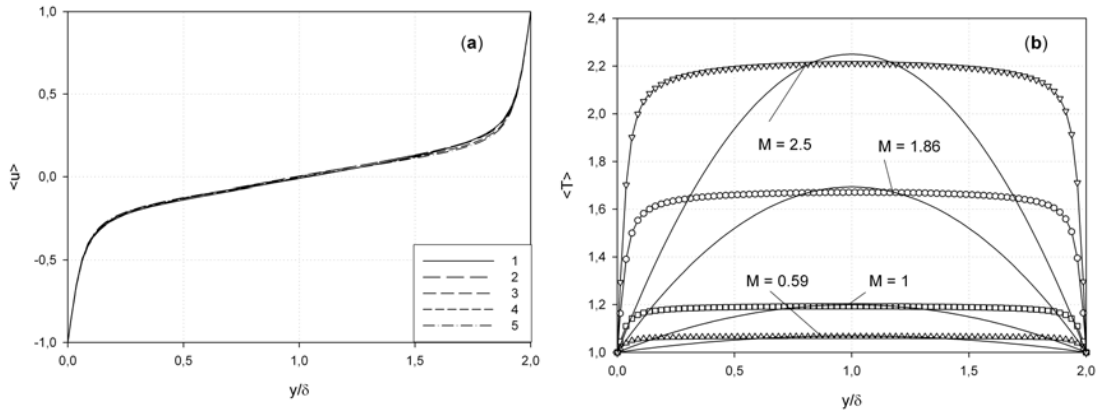

**FIGURE 5**. (a): The profiles of mean component $u$ of velocity for Couette flow at $Re_w = 3353$ and no-slip boundary conditions: 1 – incompressible flow, 2 – M = 0.59, 3 – M = 1, 4 – M = 1.86, 5 – M = 2.5. (b): The profiles of mean temperature for Couette flow at $Re_w = 3353$, different values of Mach number and no-slip boundary conditions (symbols – turbulent flow, lines – laminar flow).

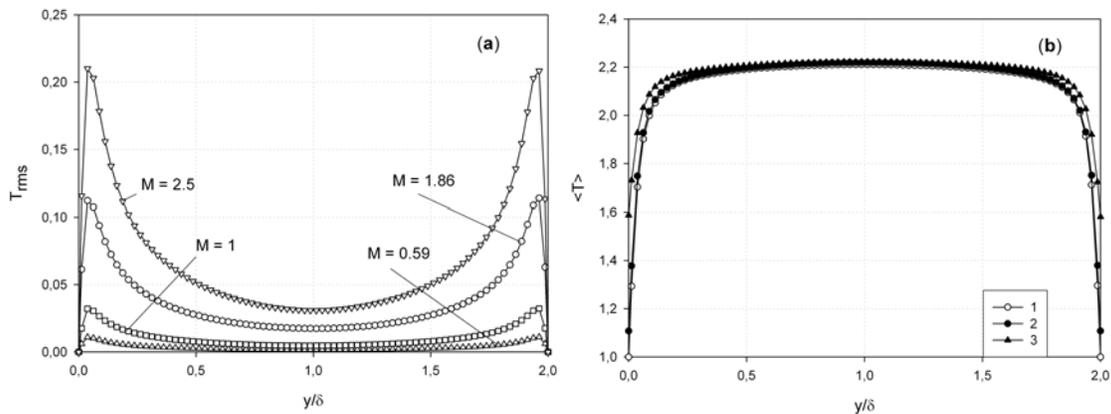

**FIGURE 6**. (a): The profiles of temperature pulsation for Couette flow at $Re_w = 3353$ and different values of Mach number for no-slip boundary conditions. (b): The profiles of mean temperature for Couette flow at $Re_w = 3353$, M = 2.5 and different values of accommodation coefficients: 1 – no-slip, 2 – $\alpha_u = \alpha_T = 0.1$, 3 – $\alpha_u = \alpha_T = 0.01$.

Therefore, the shown above results confirm that quite acceptable accuracy of description of the main characteristics of the turbulent channel flow can be achieved by using the computational domain of rather moderate size and the uniform grid over the coordinate normal to the surface. The good quality of the description of the studied flows by the developed approach seem to be provided by the following three factors:
1. The properties of the staggered grid providing the distance from the surface to the first point where the velocities components $u$ and $w$ are defined being the half of the mesh size.
2. The conservative nature of the algorithm providing the exact conservation of mass, momentum and energy for the whole flow.
3. Close to second order of approximation over time and exactly second order of approximation over spatial variables providing negligibly small effects of implicit artificial viscosity.

Let's consider the results for compressible Couette flow. The Couette flow instead of the Poiseuille flow was chosen because in this kind of flow the Mach number may be arbitrary high, while for the pressure driven Poiseuille flow the range of Mach number variation is limited.

Figure 5(a) illustrates the profiles of mean component $<u>$ of velocity for four values of Mach number. The profile of mean velocity in incompressible Couette flow is also plotted in Figure 5(a) for comparison. As it is seen, all five curves are practically merged into single curve. The profile of mean velocity $<u>$ in turbulent Couette flow of compressible gas is conservative, therefore, to the Mach number least up to $M \leq 2.5$.

The mean temperature profiles for considered values of the Mach number and no-slip boundary conditions are illustrated in Figure 5(b), the data obtained for laminar flow are also shown for comparison (it should be noted that the solution for laminar flow does not depend on the Reynolds number). As it is seen, the effect of Mach number on turbulent Couette flow of compressible gas reveals mainly in the mean temperature profiles (and hence in the mean density profiles) – in the middle of channel the temperature reaches its maximum value, the latter being increased with increasing the Mach number.

Figure 6(a) illustrates the profiles of rms temperature pulsations ($T_{rms}$) for the same variants. As it is seen, the level of pulsation is increased with increasing the Mach number. The maximum level of pulsation takes place near the surfaces as for $u$ component of velocity (see Figure 4(a)).

The profiles obtained for boundary conditions with accounting for slip effects with $\alpha_u = 1$ and $\alpha_T = 1$ were turned out practically identical to those shown. The latter is not surprising since the mean values of velocity slip $\Delta U/U_w$ for considered variants were found to be in the range $4.1 \times 10^{-3} \div 2.8 \times 10^{-3}$, while the mean values of temperature jump $\Delta T_w/T_w$ – in the range $6.6 \times 10^{-4} \div 4.5 \times 10^{-3}$.

Thus, in the considered conditions the slip effects on the compressible turbulent channel flow field are really negligibly small at the values of accommodation coefficients of the order of unity, though their values may be higher with increasing the Mach number, decreasing the Reynolds number and also with decreasing the accommodation coefficients.

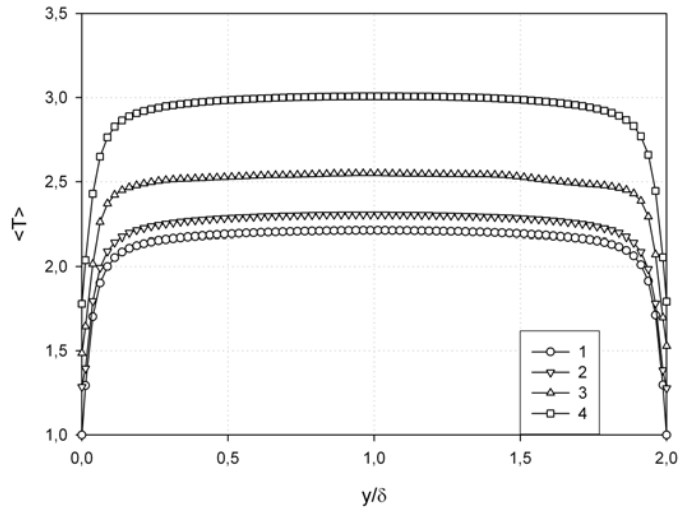

**FIGURE 7.** The profiles of mean temperature for Couette flow at $Re_w = 3353$, $M = 2.5$ and different combinations of unequal accommodation coefficients: 1 – no-slip, 2 – $\alpha_u = 1$, $\alpha_T = 0.1$, 3 – $\alpha_u = 1$, $\alpha_T = 0.03$, 4 – $\alpha_u = 1$, $\alpha_T = 0.01$.

To clarify the possible effects of small values of accommodation coefficients $α_u$ and $α_T$ their variation in the range 1 ÷ 0.01 was performed. The profiles of mean temperature for $α_u = α_T = 0.1$ and $α_u = α_T = 0.01$ together with the profile obtained for no-slip boundary conditions are illustrated in Figure 6(b). As it is seen, the small values of $α_u$ and $α_T$ lead to sufficient difference $ΔT_w$ between the surface temperature $T_w$ and the gas temperature at the surface, but in the main flow field the temperature practically does not depend on the accommodation coefficients.

The analysis of the experimental data concerning the accommodation coefficients [7] shows that the value of $α_u$ is of the order of unity practically for all combinations of flowing gas, surface and adsorbing gas. The values of $α_T$, on the contrary, may be very small, – thus for accommodation of helium on tungsten the value of $α_T = 0.0109$ at temperature -196 C [7]. That is why the variants with unequal values of $α_u$ and $α_T$ were calculated.

Figure 7 illustrates the profiles of mean temperature for Couette flow at $Re_w = 3353$, M = 2.5 and different combinations of unequal accommodation coefficients. The considerable increase of mean temperature with decreasing the accommodation coefficient $α_T$ for fixed value of $α_u = 1$ is highly unexpected, especially having in mind that for $α_u = α_T = 0.1$ and $α_u = α_T = 0.01$ no similar effects were observed (see Figure 6(b)). The discovered effect may be important in the problems of heat exchange in compressible turbulent boundary layer for some combinations of flowing gas, surface and adsorbing gas.

## CONCLUSION

The main results of the performed study may be summarized as follows:

The direct numerical simulation of compressible fully developed turbulent Couette flow between two parallel plates with temperature $T_w$ moving with velocities $±U_w$ was performed. The algorithm was tested on well known numerical solution for incompressible Poiseuille channel flow and found to provide its well description.

The profile of mean velocity $< u >$ in turbulent Couette flow of compressible gas is conservative to the Mach number at least up to M ≤ 2.5.

The effect of Mach number on turbulent Couette flow of compressible gas reveals mainly in the mean temperature profiles (and hence in the mean density profiles) – in the middle of the channel the temperature reaches its maximum value, the latter being increased with increasing the Mach number.

The slip effects in compressible turbulent Couette flow are found to be negligibly small at the values of accommodation coefficients of the order of unity. The considerable increase of mean temperature with decreasing the accommodation coefficient $α_T$ for fixed value of $α_u = 1$ was discovered. The effect may be important in the problems of heat exchange in compressible turbulent boundary layer for some combinations of flowing gas, surface and adsorbing gas.

## ACKNOWLEDGMENTS

The financial support of this study from the Siberian Branch of the Russian Academy of Sciences through the Integration Grant No 47 is gratefully acknowledged.